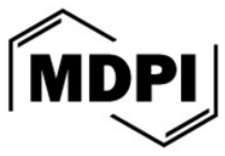

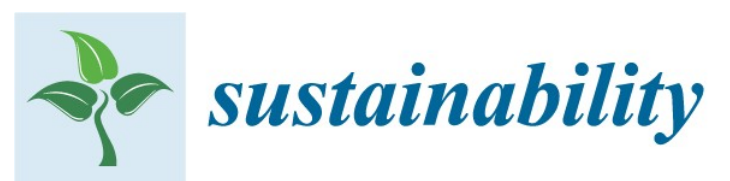

*Article*

# Combined Radar and Magnetometer Sensor Network with LoRa-Mediated Awareness for Wildlife-Vehicle Collision Prevention: A Monte Carlo Analysis

**Sergii Makovetskyi [1] and Lars Thomsen [2,*]**

[1] Kharkiv National University of Radio Electronics, Kharkiv, Ukraine; sergii.makovetskyi@nure.ua
[2] GNACODE Inc., Medicine Hat, AB, Canada; lars@gnacode.com
[*] Correspondence: lars@gnacode.com

**Abstract:** Wildlife-vehicle collisions (WVCs) cause approximately 570 human fatalities in Canada per 20-year cohort, with Alberta accounting for 22% of these and incurring an estimated CAD $300,000 per day in direct and indirect costs. Wildlife fencing combined with crossing structures reduces collisions by ~86% on well-instrumented sites but remains economically infeasible across the majority of rural road kilometres, leaving a substantial collision residual. We present a combined sensor network integrating alternating-side radar nodes (10-m spacing baseline), three-axis magnetometers, dynamic message signs, and LoRa-mediated awareness propagation between adjacent radars. System performance is evaluated through a discrete-time Monte Carlo simulation on a 1 km test corridor, incorporating a six-state animal behavioural Markov model with vehicle-threat-dependent decision branching, Intelligent Driver Model vehicle dynamics, and a three-mode contrast that isolates the contributions of sensing, driver alerting, and network coordination. Across 60 independent trials, the integrated system reduces the collision rate per road entry by 47.4% relative to an unmitigated control (Welch's $t = 2.82$, $p < 0.01$), and simultaneously increases safe road-crossing throughput by 77% by lowering the perceived vehicle threat that otherwise triggers pre-crossing retreats. Sensitivity sweeps establish a statistically significant equivalent-performance band across 5-20 m alternating radar spacing and across small-to-medium animal classes (fox- through deer-class), with operational robustness against tenfold degradation of baseline sensor sensitivity. A conservative 20 m alternating deployment spacing is recommended to provide engineering margin against range-dependent radar SNR, clutter, and environmental factors not captured in the idealized detection model. The architecture complements existing fence-and-crossing infrastructure at approximately one order of magnitude lower per-kilometre cost.

## 1. Introduction

Wildlife-vehicle collisions (WVCs) represent one of the most significant and least-mitigated road-safety hazards in Canada. Between 2000 and 2020, the Traffic Injury Research Foundation documented 570 human fatalities arising from collisions with wildlife on Canadian roads, with Alberta alone accounting for 124 of these, 22% of the national total, more than any other province or territory [1]. The Government of Alberta estimates that animal-vehicle collisions cost provincial residents approximately CAD $300,000 per day in direct property damage, emergency response, lost productivity, and human injury [2], and reports that AVCs account for roughly 60% of all reported collisions on rural Alberta highways. The scale is comparable across other developed jurisdictions: large mammal-vehicle collisions in the United States are estimated to cause over 200 human fatalities and 29,000 human injuries per year, with annual associated costs of USD 6-12 billion [3]. In Europe, ungulate collisions cause approximately 300 human fatalities and 30,000 injuries annually [3]. Beyond the human toll, WVCs are a significant socioeconomic cost driver, with reduced collisions identified as a quantifiable ecosystem service of large carnivores [4] and collision density strongly modulated by ungulate population abundance, road density, and habitat fragmentation [5].

The road-ecology literature has converged on a clear hierarchy of WVC mitigation effectiveness. Wildlife exclusion fencing combined with dedicated crossing structures (underpasses, overpasses, and jump-outs) is the most effective intervention available, reducing large-ungulate collisions by approximately 86% on well-instrumented sites. The standardised assumption is adopted in the cost-benefit decision-support model of Huijser et al. [6] and consistent with Rytwinski et al.'s meta-analytic finding that fencing combined with crossing structures reduced large-mammal road-kill by 83% [7]. Site-specific reductions in the 82-96% range have been documented along the Trans-Canada Highway in the Canadian Rocky Mountains, with apparent hoofed animal reductions of up to 96% [8] and realized effectiveness of 82-90% after accounting for background trends at unmitigated control sites [8], findings further corroborated by long-term reported-WVC analyses in the Bow Valley [9]. However, these reductions are conditional on fence sections of at least 5 km, regular maintenance, and the presence of crossing structures; shorter fence segments are substantially less effective, and gaps developing through long-term degradation reintroduce collisions [10,11]. At best-instrumented sites, approximately 4-19% of collisions persist near fence-ends and along adjacent unmitigated road sections [8,12], with Big Pine Key specifically showing a 95% reduction in the mitigated section accompanied by significant hotspot formation at the fence-end and on adjacent unmitigated highway [12]. Fence-and-crossing-structure systems require substantial up-front capital that is rarely justified outside collision hotspots. Alberta's Bow Valley Gap overpass (completed 2024) cost CAD $17.5 million with 12 km of associated wildlife fencing which is approximately CAD $1.46 million per kilometre of treated corridor [15] and was sited where annual collision damages on the Trans-Canada Highway exceeded CAD $750,000 per year. On the much larger fraction of rural road network where WVC damage is substantial in aggregate but not spatially concentrated, this cost profile is prohibitive. Animal detection systems (ADS) offer a complementary mitigation pathway at materially lower cost, with meta-analytic effectiveness of 57% [7] and individual deployments reporting reductions of up to 75% in first-year evaluations [13].

This paper presents a combined radar and magnetometer sensor network with LoRa-mediated awareness propagation, designed to address the residual WVC burden at fenced sites and the much larger gap along the unfenced rural road network. The system integrates alternating-side radar nodes, three-axis magnetometers, driver-facing dynamic message signs, and a network coordination layer in which detection events on one radar boost the sensitivity of adjacent radars over a short persistence window. We evaluate the architecture using a discrete-time Monte Carlo simulation incorporating a six-state behavioural Markov model for animal motion and the Intelligent Driver Model for vehicle dynamics, with a three-mode experimental design that cleanly isolates the contributions of sensing, driver alerting, and network coordination. The paper reports headline collision-reduction performance, three sensitivity analyses establishing the operational envelope, and a cost-benefit positioning of the proposed system relative to existing infrastructure.

## 2. Materials and Methods

### *2.1. System Architecture*

The proposed sensor network is illustrated schematically in Figure 1a. Radar nodes are placed along the road shoulder at a baseline spacing of 20 m. Each radar has a nominal detection radius of 15 m, oriented to monitor the wildlife approach zone adjacent to the carriageway. The geometry is chosen such that the worst-case detection distance to any point in the immediate roadside zone satisfies $R_{det} \geq \sqrt{(s^2 + d_y^2)}$, where s is the radar spacing and $d_y$ is the maximum animal distance to be covered, ensuring no gap in coverage at the deployed baseline parameters.

Three-axis magnetometer sensors are interleaved with the radars at 200 m intervals, providing a secondary detection modality based on perturbations of the local geomagnetic field by ferromagnetic material in the vicinity. The magnetometers are non-specific with respect to source material and respond to any object capable of distorting the local field, with signal amplitude scaling with the ferromagnetic mass; larger vehicles therefore produce stronger signatures than smaller vehicles. In the present work, magnetometer data are used to support vehicle-state estimation that is fused with the radar detection stream at the network coordination layer.

Driver-facing dynamic message signs (DMS) are placed at the corridor edges and at intermediate locations to deliver the alert payload. All sensor nodes communicate over a hybrid backbone: low-voltage differential signalling over RJ50 for the closely-spaced radar-to-radar links, and LoRa (sub-GHz, low-power) for the longer-range network coordination layer.

The network coordination mechanism operates as follows. When any radar detects an animal, the detecting node broadcasts a notification over the LoRa channel to all radars within a 1500 m awareness range. Each receiving radar applies a sensitivity boost factor of 1.8× to its per-frame detection rate for a persistence window of 30 s following the most recent broadcast, after which the boost decays back to baseline. The driver-facing DMS is activated by any successful detection regardless of source and remains active for the same 30 s persistence window or until no detected animal remains in a dangerous behavioural state, whichever is later. This decoupling of *driver-alerting* from *sensor coordination* is central to the experimental design described in Section 2.4.

The LoRa backbone is divided into two sub-channels operating on separate sub-band frequencies: an internal sensor-mesh *awareness channel* carrying detection notifications between radar nodes (radar-to-radar mesh, 1500 m range, short bursty broadcasts), and an external *alert channel* carrying caution-speed and warning payloads from the edge coordinator to DMS units and to optional in-vehicle receivers. The two channels can be served by a single dual-band-capable LoRa transceiver chip (for example the Semtech SX1262 family) operating on two sub-band allocations under software control, or by two physical radios where simultaneous transmit-receive across both channels is required. The motivation for this separation, which trades the simpler single-channel design for airtime, security, and forward-compatibility benefits, is discussed in Section 4.2.

*2.2. Simulation Framework*

System performance is evaluated using a discrete-time Monte Carlo simulator implemented in Python. The simulation runs at a fixed time step of Δt = 0.1 s on a 1000 m test corridor with four vehicles per direction. The full simulator source code is released open-source as supplementary material to ensure reproducibility (see Section *Data Availability Statement*) and see Makovetskyi and Thomsen for hardware details [16,17].

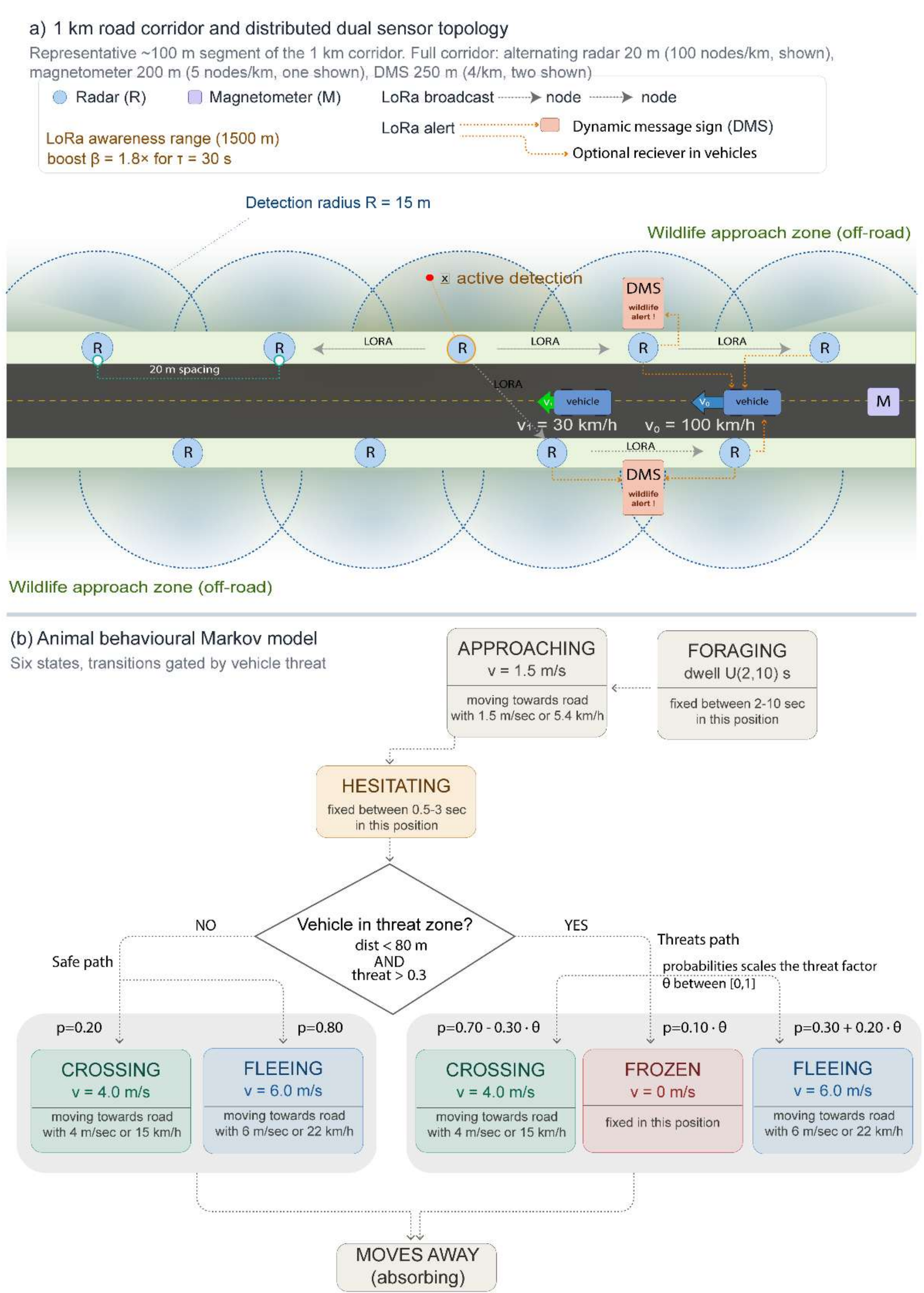

**Figure 1.** System architecture and behavioural models. **(a)** Top-view sensor topology along a representative 1 km road corridor showing one active detection event. Alternating radar nodes are deployed at 20 m spacing (50 nodes/km), three-axis magnetometer sites at 200 m intervals (5/km), and driver-facing dynamic message signs at 250 m intervals (4/km). When any radar detects an animal it broadcasts an "awareness" notification over LoRa to all radars within a 1500 m range, applying a sensitivity boost $\beta = 1.8\times$ for $\tau = 30$ s; the central DMS simultaneously switches to caution-speed display (→30 km/h). The dashed circle shows the 15 m detection radius around the active radar. **(b)** Six-state animal behavioural Markov model. Transition probabilities depend on the proximate vehicle situation: the vehicle-threat branch (HESITATING → FROZEN or FLEEING); or the no-threat branch (CROSSING or FLEEING ). Reduced cruising speed under driver alert shifts probability mass from the vehicle-threat branches toward the no-threat CROSSING transition.

#### 2.2.1. Vehicle Dynamics

All vehicles follow the Intelligent Driver Model (IDM) of Treiber, Hennecke, and Helbing [14]. At each time step, vehicle acceleration is computed as:

(1) $$a = a_{max} [1 - (v / v_0)^{\delta} - (s^*(v, \Delta v) / s)^2], \text{ Equation 1}$$

desired gap distance to the vehicle ahead:

$$s^* = s_0 + vT + v\Delta v / (2\sqrt{(a_{max} b_{comf})}), \text{ Equation 2}$$

parameters $s_0 = 5$ m, $T = 1.5$ s, $a_{max} = 2.5$ m/s$^2$, $b_{comf} = 4.0$ m/s$^2$, $\delta = 4$, and emergency braking up to $a_{em} = 9.0$ m/s$^2$ when a driver-alert is active.
The baseline desired speed is $v_0 = 100$ km/h (cruise), reduced to 30 km/h (caution) when the driver is alerted. Driver perception-reaction time is fixed at 1.5 s, in line with the AASHTO highway design standard.

#### 2.2.2. Animal Arrival and Behaviour

Animals arrive at the corridor according to a homogeneous Poisson process with rate $\lambda = 15$ /hour, spawned uniformly along the road length at y = -25 m (off-road side). Each spawned animal is assigned a body-size factor sampled from a three-component mixture: small (15%, $\sigma_{RCS} \in [0.25, 0.55]$, fox/coyote class), medium (60%, [0.7, 1.2], deer class), and large (25%, [1.4, 2.3], elk/moose class). Animals progress through the six-state behavioural Markov model shown in Figure 1b: FORAGING → APPROACHING → HESITATING → {CROSSING, FROZEN, FLEEING} → MOVES AWAY.

The FORAGING state has random dwell uniform in [2 s, 10 s]; HESITATING dwell is uniform in [0.5 s, 3 s]. From HESITATING, transition probabilities depend on the proximate vehicle situation: when no vehicle is present in a threat zone, p(CROSSING) = 0.80; when a vehicle is present, p(FROZEN) = 0.10, p(FLEEING) = 0.20. During a CROSSING traversal, a base freeze probability of 0.15 per dangerous interaction is applied if the animal perceives an emergency-braking vehicle. Approach, crossing, and flight speeds are 1.5, 4.0, and 6.0 m/s respectively, consistent with reported field values for medium and large ungulates.

#### 2.2.3. Radar Detection Model

At each time step, every animal within radar range is evaluated for first-time detection. The per-frame detection probability is modelled as:

(2) $$p_{detect} = 1 - \exp(-\kappa \cdot f_{size}(\sigma) \cdot \beta \cdot \Delta t) \text{ Equation 3}$$

where $\kappa$ is the baseline per-second detection rate (default $\kappa = 3.0$ s$^{-1}$ for a reference $\sigma_{RCS} = 1$ target), $f_{size}(\sigma) = \min(2.0, 0.4 + 0.6\sigma)$ scales sensitivity with radar cross-section, and $\beta \in \{1.0,$ 1.8} is the neighbour-sensor sensitivity boost applied during awareness propagation. This exponential per-frame model is consistent with classical radar detection theory under multi-pulse non-coherent integration. Once an animal is detected, it remains detected for the remainder of its corridor traversal; subsequent radar passes do not generate additional events.

### *2.3. Three-Mode Experimental Design*

The simulator can be executed in three operating modes, the differences between which constitute the core of the experimental design:

1. **Control.** Sensors and DMS are absent. Drivers operate at cruise speed throughout. This is the unmitigated baseline.
2. **Detection.** Sensors are active and emit driver alerts via DMS on every successful detection. Drivers reduce to caution speed for the awareness persistence window. There is *no* propagation of sensitivity boost between radar nodes.
3. **Aware.** Identical to Detection, with the addition of LoRa-mediated sensitivity-boost propagation: a detection at one radar boosts neighbouring radars by a factor $\beta = 1.8$ for 30 s.

This three-mode contrast cleanly separates the contributions of (i) sensing combined with driver alerting (Control vs Detection) and (ii) network coordination (Detection vs Aware). Two-mode designs reported in the prior ADS literature merge these two contributions and cannot attribute reductions to the responsible mechanism.

### *2.4. Sensitivity Sweeps*

In addition to the headline three-mode comparison, three one-dimensional sensitivity sweeps were conducted to characterise the system's operating envelope. Each sweep varies one parameter, radar spacing $s \in \{5, 10, 15, 20, 25, 30, 40\}$ m, animal size scaling $\sigma_{scale} \in \{0.25, 0.5, 0.75, 1.0, 1.5, 2.0, 3.0\}$, or baseline detection rate $\kappa \in \{0.3, 0.5, 1.0, 2.0, 3.0, 5.0\}$ $s^{-1}$, holding all other parameters at their default values. The size sweep applies a multiplicative scaling to all sampled animal sizes; the detection-rate sweep models degraded sensor sensitivity from environmental factors such as precipitation, foliage occlusion, or hardware aging.

### *2.5. Trial Counts and Statistical Analysis*

The headline comparison comprises 20 independent trials of 4 simulated hours per mode (60 total trials). Each sensitivity sweep comprises 15 trials of 2 simulated hours per (mode, parameter value) pair, giving 270-315 trials per sweep. All trials use independent random seeds. Per-trial outcomes (collisions, road entries, detected animals, in-range detection latency, and animal state-visit counts) are aggregated and compared across modes using two-sided Welch's t-tests for unequal variances. Reported uncertainties are sample standard deviations (Bessel-corrected, n-1 denominator). Significance is reported at the conventional thresholds $p < 0.05$ (*), $p < 0.01$ (**), and $p < 0.001$ (***).

## 3. Results

### *3.1. Headline Three-Mode Comparison*

The headline experiment comprised 20 independent trials of 4 simulated hours per mode (60 trials total), at default parameters: 15 m alternating radar spacing, $\lambda = 15$ animals/hour arrival rate, $\kappa = 3.0$ $s^{-1}$ baseline detection rate. Outcomes are summarized in Table 1.

**Table 1.** Three-mode comparison statistics, 20 trials × 4 h per mode. Reported as mean ± 1 SD.

| Metric | Control | Detection | Aware | Statistic |
|---|---|---|---|---|
| Collision rate per road entry (%) | 10.46 ± 7.07 | 6.24 ± 3.47 | 5.50 ± 3.44 | Control vs Detection: -40.4% ($t = 2.40$, $p < 0.05$); Control vs Aware: -47.4% ($t = 2.82$, $p < 0.01$) |
| Detection rate (%) | 0 | 98.81 | 98.98 | Size-invariant |
| In-range detection latency (s) | n/a | 0.294 ± 0.041 | 0.297 ± 0.049 | Within kinematic budget |
| Animals arriving per trial | 59.0 | 59.0 | 59.0 | Identical Poisson input |
| Road entries per trial | 22.9 | 41.1 | 40.6 | +77% vs Control |
| Crossing success rate (%) | 38.9 | 69.6 | 68.8 | Throughput nearly doubled |
| Cumulative frozen-on-road time (s) | 11.5 | 1.4 | 0.4 | -96% vs Control |

The collision rate per road entry falls from 10.46% under Control to 5.50% under Aware, a 47.4% relative reduction (Welch's $t = 2.82$, $df = 27.5$, $p < 0.01$). The Control-to-Detection step alone delivers a 40.4% reduction ($t = 2.40$, $df = 27.6$, $p < 0.05$), establishing the sensing-and-alerting layer as the principal contributor. Detection probability is high in both sensor-active modes (98.8% under Detection, 99.0% under Aware) with in-range latency of 294 to 297 ms, well within the kinematic stopping budget for cruise-speed vehicles.

The mechanism is straightforward. The same 59 animals per trial arrive at the corridor in all three modes, but their behaviourat the road edge differs. Under Control, cruise-speed vehicles trigger the threat-perception branch of the behavioural model, and only 38.9% of animals enter the road; the rest hesitate, flee, or freeze before crossing. Under Detection and Aware, the dynamic message signs reduce vehicle speed to caution levels during the awareness window. The slowed vehicles fall below the threat-perception threshold (Section 2.2.2), so the same animals now cross with probability of 0.80. Road entries rise to 69.6% under Detection and 68.8% under Aware, a 77% increase relative to Control. This is not a negative finding: the simulation models corridors where animals need to cross for foraging, migration, and habitat connectivity, and a system that lets them do so safely is consistent with the ecological-connectivity goals that motivate WVC mitigation. Cumulative frozen-on-road time, which captures dangerous mid-crossing immobility, falls from 11.5 s under Control to 0.4 s under Aware, a 96% reduction. Detection probability is 98.8 to 99.0% with in-range latency of 294 to 297 ms in both sensor-active modes, well within the kinematic stopping budget.

Together, these results describe a dual mitigation mechanism: (i) reduced per-crossing risk through kinematic slowing of vehicles in the alert zone, and (ii) increased crossing throughput because the slowed vehicles no longer trigger the threat-perception branch in approaching animals. Both effects contribute to the 47.4% rate reduction. The qualitative pattern is consistent with the 57% meta-analytic mean for animal detection systems reported in the road-ecology literature [7] and with the 75% first-year reduction from the most-comparable single-site deployment [13], placing this architecture squarely within the empirical effectiveness band for the ADS intervention class.

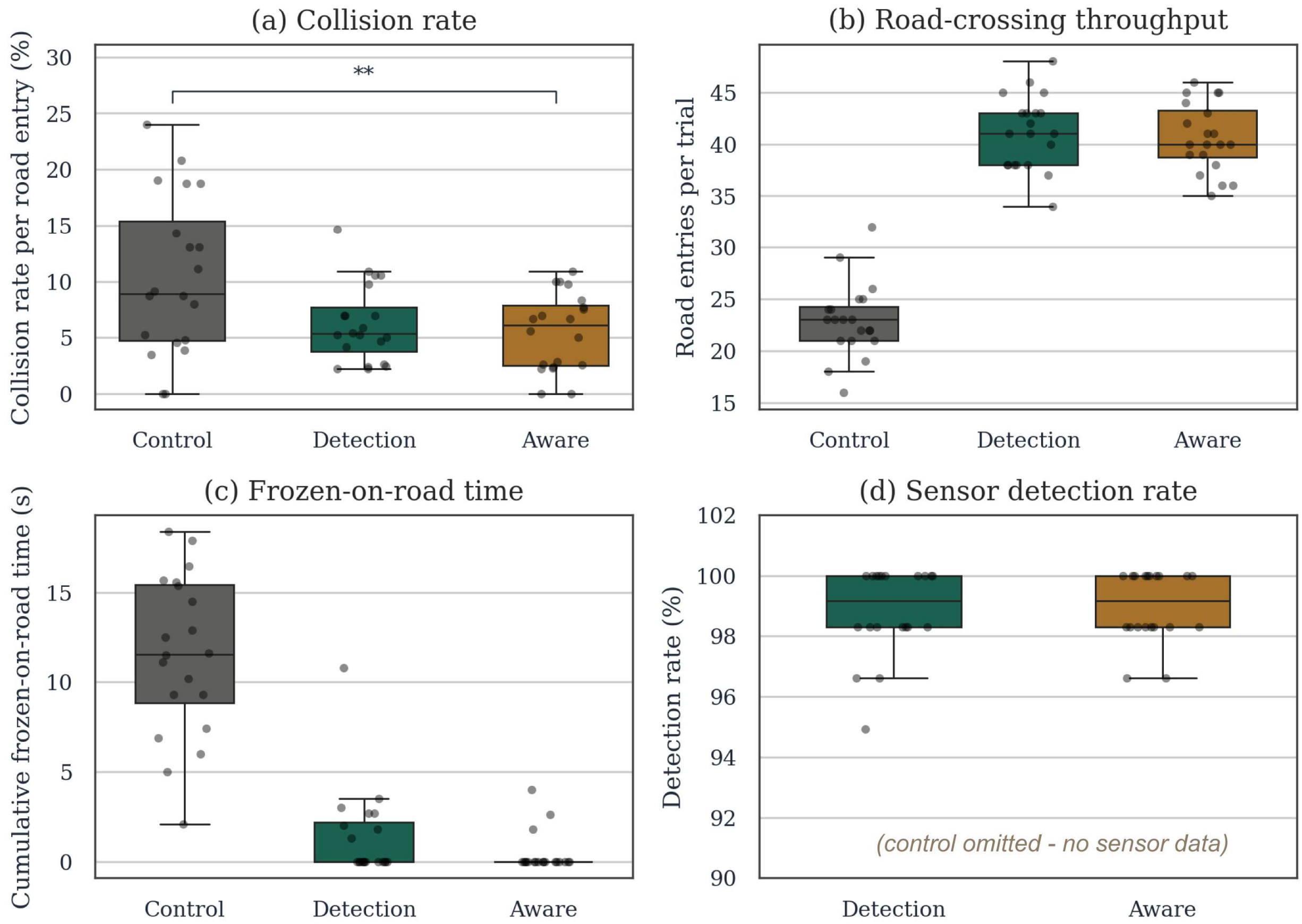

**Figure 2.** Four-panel three-mode Monte Carlo comparison (20 trials × 4 h per mode). Each point is one trial; whiskers show Tukey range. Absolute collision counts (top-left) are statistically indistinguishable across modes, but the *collision rate per road entry* (top-right) falls from 10.5% under Control to 5.5% under Aware (-47.4%, $p < 0.01$). The throughput panel (bottom-left) shows the underlying behavioural mechanism: road entries per trial rise from 22.9 to 41 in sensor-active modes, indicating that slowed vehicles allow more animals to attempt and complete crossings safely. Cumulative frozen-on-road time (bottom-right) falls by 96%. Interactive version with hover-inspectable per-trial values is available at https://gnacode.github.io/WILDLIFE-VEHICLE-COLLISION-MONTECARLO/.

*3.2. Radar Spacing Sweep*

The spacing sweep characterises the cost-performance trade-off in sensor density. Spacing values from 5 to 40 m were tested at 15 trials × 2 h per (mode, spacing) point, with all other parameters at default. The collision-rate-per-road-entry reduction (Control vs Detection) is statistically significant across the 5-20 m range and falls to non-significance from 25 m onward: 66.4% at s = 5 m ($p < 0.01$), 59.7% at 10 m ($p < 0.05$), 53.1% at 15 m ($p < 0.05$), 53.0% at 20 m ($p < 0.05$), 42.1% at 25 m (n.s.), 28.3% at 30 m (n.s.), and 23.6% at 40 m (n.s.). Detection rate falls smoothly across this range, from 100% at s = 5-10 m through 98% at 15 m and 95% at 25-30 m, dropping sharply to 70% at 40 m alternating spacing where coverage gaps become geometrically unavoidable.

This profile establishes 5-20 m alternating spacing as the equivalent-performance band: within this range, collision rates and significance thresholds are statistically indistinguishable, with the system delivering 53-66% reduction in collision rate. Beyond 20 m the reduction degrades smoothly with increasing sensor sparsity. At 40 m alternating (worst case tested) the system still delivers a 24% reduction relative to Control but is no longer statistically demonstrable at the n = 15 trials-per-point used here.

This is consistent with the coverage geometry: at 40 m alternating the worst-case midpoint distance from any radar exceeds the 15 m detection radius, producing detection gaps that the simulator's idealised model still partially recovers from via animal exposure-time saturation but with degrading performance. The simulator's detection model treats the 15 m radius as a hard cutoff at full sensitivity and does not represent range-dependent radar SNR, angular clutter, weather-induced sensitivity reductions, or aspect-angle effects encountered in operational deployment. We therefore recommend a conservative 20 m alternating spacing for operational deployment (50 radars per kilometre, 40 m same-side spacing). This is precisely the last statistically-significant point in the sweep, preserves engineering margin against RF-propagation and environmental effects omitted from the model, and delivers simulated collision reduction within 1 percentage point of denser deployments while halving the per-kilometre sensor count relative to a 10 m configuration.

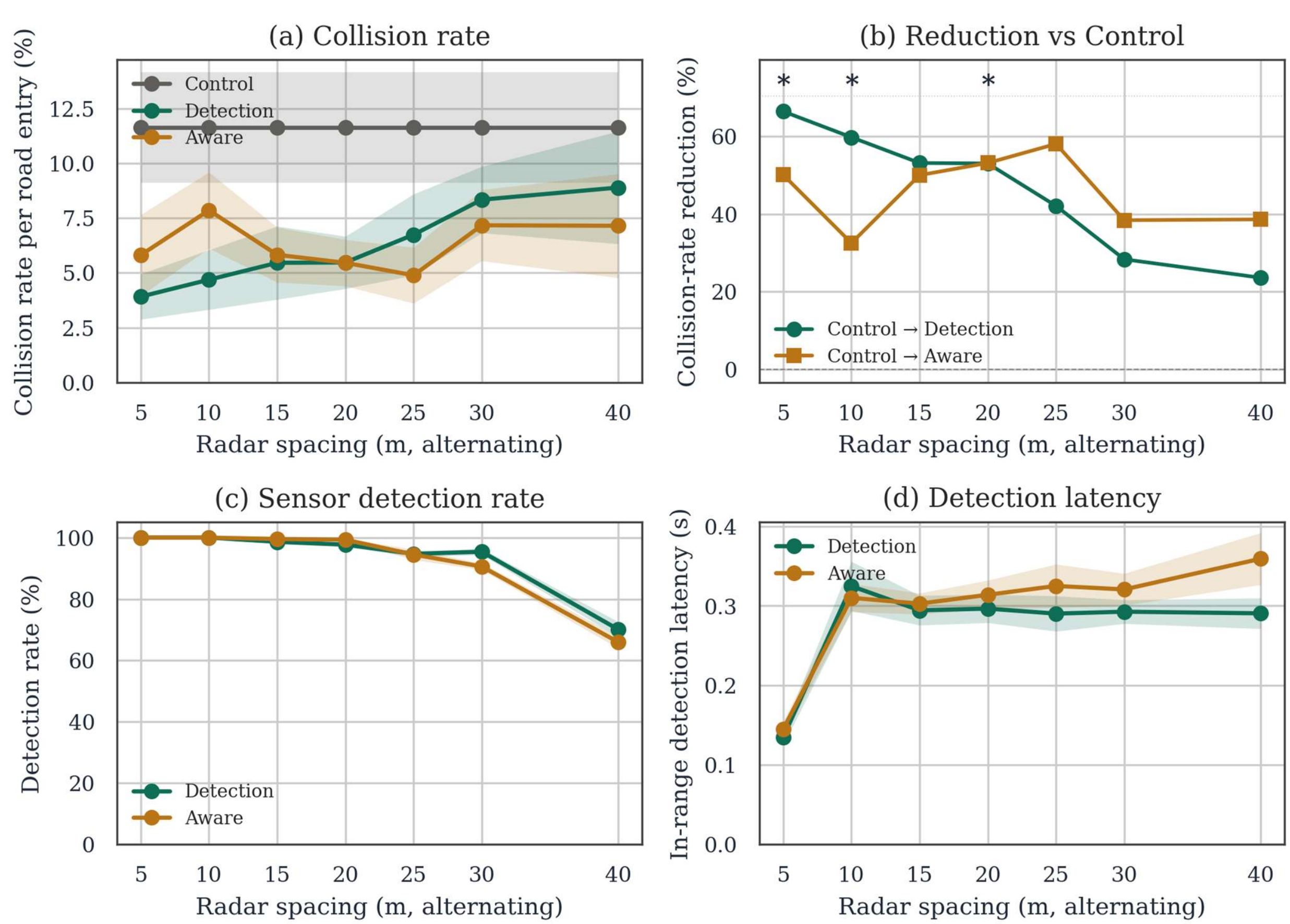

**Figure 3.** Sensitivity sweep on radar spacing. Each point is the mean of 15 independent trials of 2 h; error bars (shaded area) show ± 1 SD. The system performs equivalently across 5-20 m alternating spacing (the last statistically-significant point), with smooth degradation thereafter. The Aware-mode advantage over Detection-only emerges at sparser spacings (25-40 m) where the network coordination layer compensates for coverage gaps, although this contrast does not reach statistical significance at the trial count tested.

### *3.3. Animal Size Sweep*

Animal body size was scaled by a multiplicative factor $\sigma_{scale} \in [0.25, 3.0]$, modelling target species from fox/coyote class (0.25×) through medium deer (1.0×) up to large moose (3.0×). Detection rate is remarkably invariant across this range, remaining between 98.4% and 99.5% for every size scaling tested. In-range latency, by contrast, scales smoothly with size as predicted by the size-dependent term $f_{size}(\sigma)$ in the detection model: 0.532 s at the smallest size, 0.376 s at $\sigma_{scale} = 0.75$, 0.294 s at the default size, falling to 0.167 s at the largest size, a 3.2-fold range that nonetheless remains within the kinematic stopping budget for cruise-speed vehicles at all sizes tested.

The system delivers substantial collision-rate reductions across the entire size range tested. For fox- through deer-class targets, reductions are 56.7% at $\sigma$scale = 0.25, 55.7% at 0.50, 59.4% at 0.75, and 53.1% at 1.00 (all $p < 0.05$). For elk- and moose-class targets, reductions are 43.1% at $\sigma$scale = 1.50, 39.0% at 2.00, and 42.7% at 3.00. The sensor detection rate remains between 98% and 99% across the entire range, so heavier ungulates are detected as reliably as smaller carnivores. The narrowing of the reduction at the large-animal end reflects a modest upward trend in the Detection-mode collision rate, from 5.46% at $\sigma$scale = 1.0 to 6.63 to 7.10% at $\sigma$scale ≥ 1.5. This gradient is a property of the simulator rather than of animal detection systems in general. The behavioural Markov model (Figure 1b) and the IDM vehicle dynamics (Equation 1) are size-independent: the alerted vehicle decelerates to the same caution-speed setpoint regardless of which animal triggered the alert, and animal kinematics (approach, crossing, flight speeds) do not vary with body size. A size-aware kinematic response, in which heavier-mass detections trigger proportionally larger speed reductions, is a natural forward design direction (Section 4.3) and would be expected to flatten the residual gradient.

Independently of the simulator's size-independence, the empirical road-ecology literature reports substantially higher reductions for fencing combined with crossing structures (83% [7], up to 96% [8]) than for ADS deployments (57% [7], 75% in deer-focused single-site evaluations [13]). For corridors where elk- and moose-class species dominate the collision risk, physical exclusion therefore remains the better-evidenced primary intervention on the strength of this empirical effectiveness gap. Animal detection systems are best positioned for the much larger road-network footprint where collision risk is real but spread thinly over many kilometres for which fence-and-crossing infrastructure is economically out of reach.

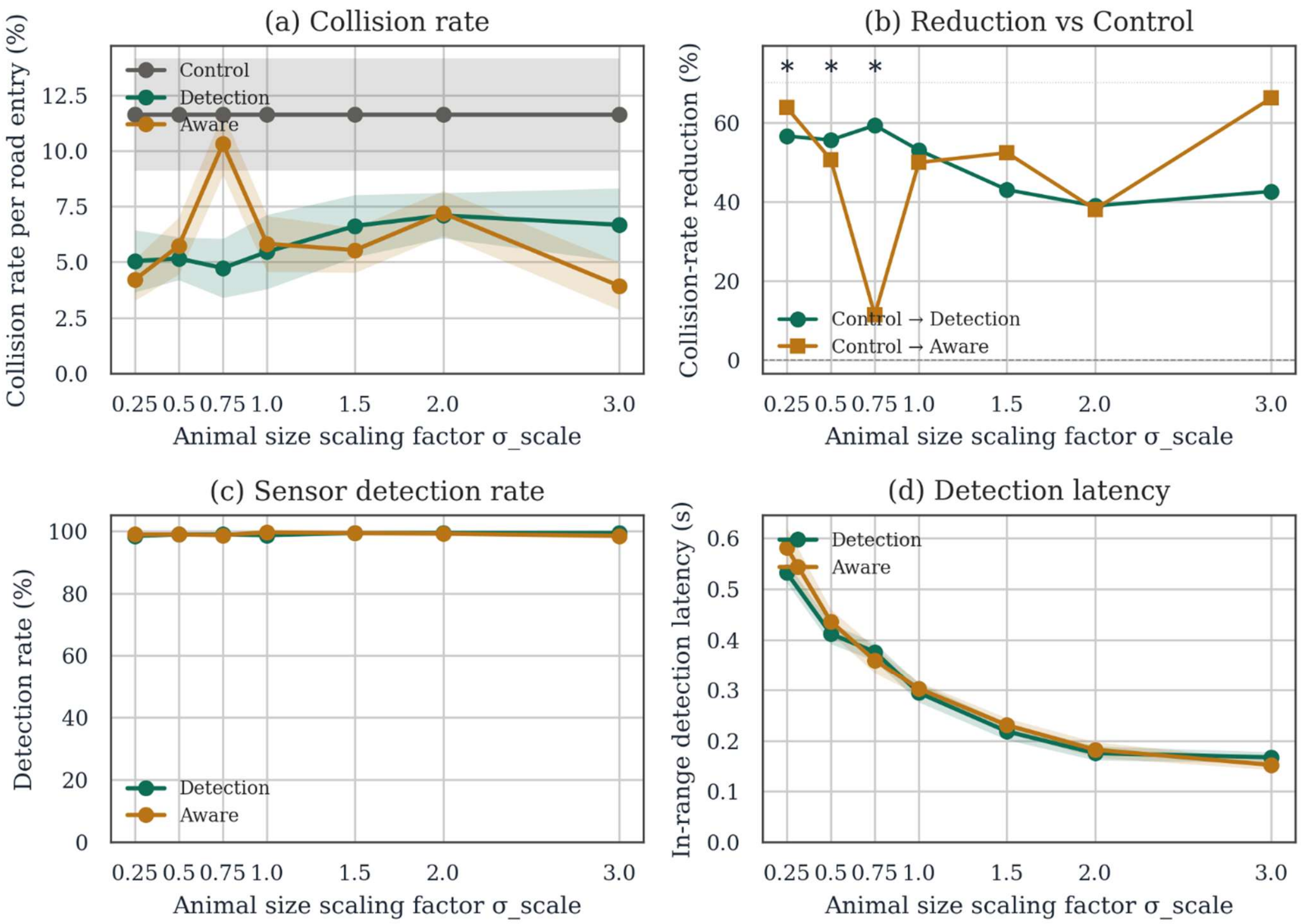

**Figure 4.** Sensitivity sweep on animal size scaling factor $\sigma_{scale}$. Detection rate is size-invariant at 98-99% across the full range. Collision-rate reduction relative to Control is statistically significant for small-to-medium animals ($\sigma_{scale} \leq 1.0$) and trends downward for elk- and moose-class animals, consistent with the size-independent vehicle response in the simulator. In-range detection latency scales inversely with size as predicted by the radar detection model, but remains within the kinematic stopping budget at all sizes tested.

### *3.4. Baseline Sensor Sensitivity Sweep*

The mechanism underlying this robustness is the long animal exposure time within radar coverage. Animals spend 2-10 s in foraging followed by ~5 s in approach within sensor range; the cumulative detection probability $P = 1 - \exp(-\kappa \cdot t_{exposure})$ saturates rapidly. At $\kappa = 0.30\ s^{-1}$ and $t_{exposure} = 10$ s the cumulative probability is $1 - e^{-3} = 95.0\%$, consistent with the observed detection rate of 95.2% at this point. The practical implication is that the system tolerates substantial degradation in per-frame sensor sensitivity: a 10x reduction from the default κ leaves the collision-reduction effectiveness essentially intact. This is the most operationally important finding of the sensitivity sweep: weather-degraded sensors, fouled antennas, or sub-optimally-tuned hardware all degrade the per-frame detection probability but do not appreciably degrade the system's end-to-end protective effect, because the multi-second animal exposure time compensates by integration.

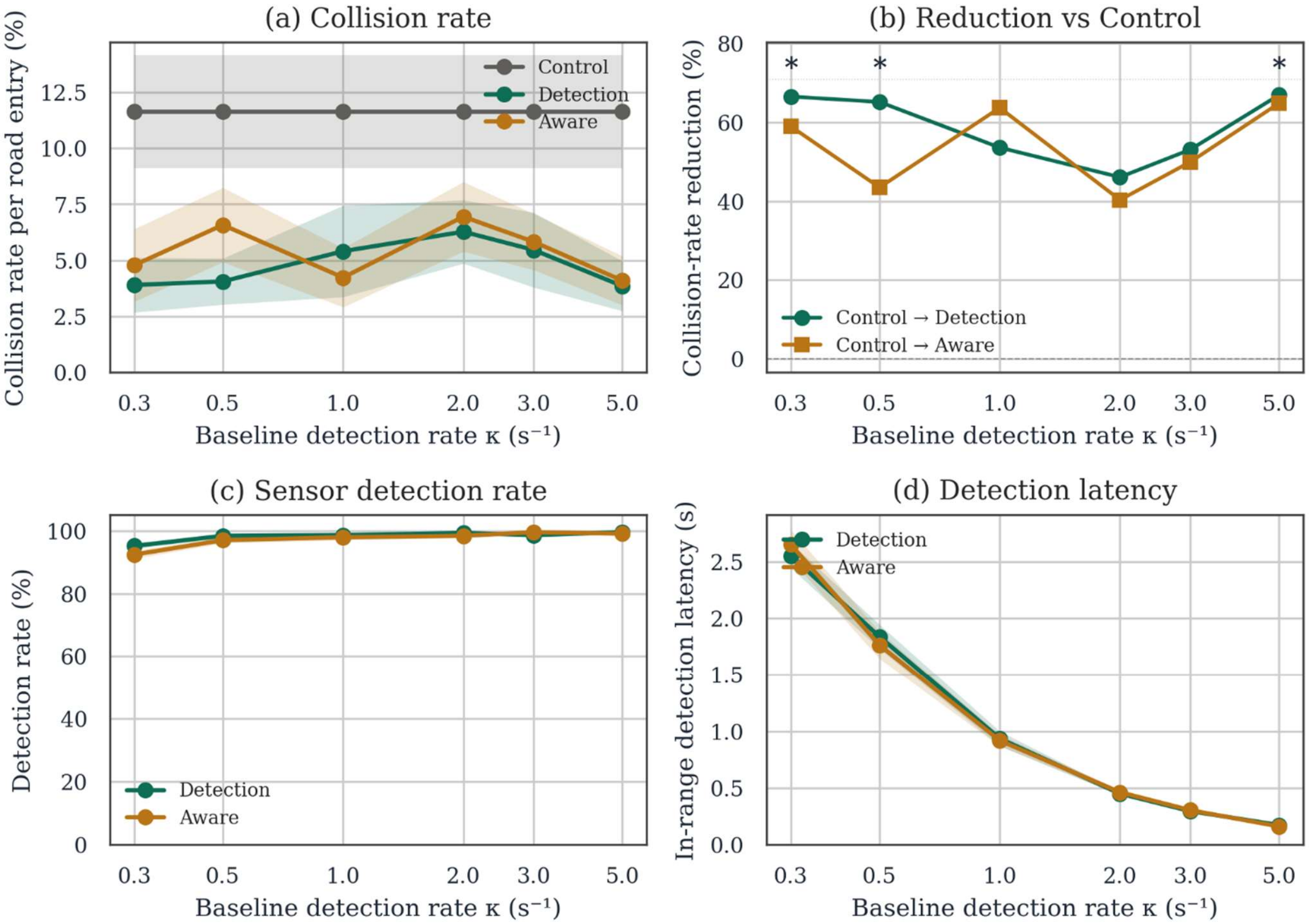

**Figure 5.** Sensitivity sweep on baseline per-second detection rate κ. The system retains 46-67% collision reduction across the entire range, with detection rate ≥ 95% even at a tenfold reduction in κ from the default. Statistical significance is recovered at both extremes of the sweep; the loss of significance in the middle ($\kappa = 1\text{-}2\ s^{-1}$) reflects per-trial variance rather than a mechanism-level performance gap, with the underlying mean collision rate varying by only a few percentage points across the full sweep.

## 4. Discussion

### *4.1. Cost-Benefit Positioning*

The estimated per-kilometre capital cost of the proposed radar-magnetometer-DMS sensor network is approximately CAD $80,000-160,000, comprising 50 radar nodes, 5 magnetometer sites, 4-6 dynamic message signs, edge-computing nodes, backbone infrastructure, off-grid power systems, and installation labour. Annual maintenance is estimated at 10-15% of the capital cost. The system is therefore approximately one order of magnitude less expensive per treated kilometre than fence-and-crossing infrastructure (CAD $100,000 vs CAD $1,460,000).

This cost ratio does not imply that the sensor network is the preferred intervention everywhere. Where annualised collision damages are high relative to the per-kilometre cost, for instance, the Bow Valley Gap section of Highway 1, where pre-mitigation collision damages reached approximately CAD $750,000 per year [9], yielding a fence-and-overpass payback period of approximately 23 years considering property damage alone, comprehensive fence-and-crossing infrastructure delivers superior collision reduction and the additional ecological benefits of preserved habitat connectivity. The sensor network's positioning is instead the much larger road-network footprint where annualised damages of CAD $5,000-50,000 per kilometre are insufficient to amortise a CAD $1.5 million/km capital investment but more than sufficient to amortise a CAD $100,000/km system over a 5-10 year operational life. This positioning makes layered, low-cost detection-and-alert systems an economically defensible primary mitigation on rural road kilometres for which fence-and-crossing infrastructure is presently out of reach.

### *4.2. Alert-Channel Architecture*

#### 4.2.1. Two-Channel Network Architecture

As introduced in Section 2.1, the architecture distinguishes between an internal *awareness channel* (radar-to-radar mesh broadcast over LoRa, 1500 m range, 30 s persistence) and an external *alert channel* (edge-coordinator-to-DMS and optionally edge-coordinator-to-vehicle delivery, on a separate LoRa sub-band).

#### 4.2.2. Driver-Facing Alert Channels

The simulation uses a driver-facing dynamic message sign (DMS) as the alert delivery channel, since roadside variable-message and active-warning signs are the dominant alert mechanism in deployed ADS installations and require no driver-side equipment. The sensor network architecture does not depend on this choice: alternative or complementary channels can be substituted or layered without redesign of the sensing or coordination layers

### *4.3. Limitations and Future Work*

The present analysis is in-silico. The underlying radar nodes and IoT-mesh signal-processing chain have been independently developed and characterised in adjacent applications [16,17], but the sensor combination, network-coordination layer, and behavioural-Markov evaluation framework presented in this paper are new. Field validation under operational conditions (including driver compliance with DMS-displayed speed reductions, sensor performance across seasonal weather variations, and longitudinal collision-rate measurement against a control corridor) is the necessary next step before quantitative deployment recommendations can be transferred to operational specifications. The behavioural Markov model used here captures the principal animal-state transitions identified in the field-ecology literature but does not differentiate by species, time of day, or seasonal-migration phase; species-specific parameterisation would improve simulation fidelity for corridors where one ungulate species dominates the collision risk.

The radar detection model does not consider several effects relevant to operational deployment. The 15 m detection radius is treated as a hard cutoff at full sensitivity within range and zero detection beyond, whereas real radar detection probability falls smoothly with range as approximately $R^{-4}$ (the monostatic radar equation under non-coherent integration). The model also omits angular clutter, weather-induced attenuation (rain, snow), foliage occlusion, and aspect-angle dependence of the radar cross-section. These omissions are appropriate for the architectural-comparison purpose of the present study, where the relevant question is which combinations of sensing and coordination yield the largest collision reductions, but they mean that the simulated 5-25 m equivalent-performance range overestimates the operational tolerance to wider deployment spacings. The conservative 20 m alternating recommendation in Section 3.2 reflects this consideration; operators with site-specific RF propagation measurements may justify tighter or looser spacings around this nominal value.

Three areas of forward technical work are particularly attractive. First, fusion of the radar and magnetometer streams at the network coordination layer offers improved discrimination of approaching vehicles from approaching animals; this refinement could substantially reduce DMS false-alarm rates and improve long-term driver-compliance behaviour. Second, integration with existing fence systems at fence-end zones, where collisions concentrate after fencing is deployed, offers a high-value combined-architecture deployment with quantifiable additive benefit. Third, emerging vehicle-to-infrastructure communication standards provide a forward-compatible evolution path for the alert delivery layer; integrating these channels with the sensor network is a substantial separate engineering program but justifies the present sensor architecture's continued relevance over a 5 to 10 year deployment horizon.

## 5. Conclusions

This paper presents a combined radar-magnetometer sensor network with LoRa-mediated awareness for the prevention of wildlife-vehicle collisions on rural transport corridors. Discrete-time Monte Carlo simulation incorporating a vehicle-threat-dependent behavioural animal Markov model and Intelligent Driver Model vehicle dynamics demonstrates a 47.4% reduction in collision rate per road entry relative to an unmitigated control ($p < 0.01$) across 60 independent trials, achieved through a dual mechanism: kinematic slowing of alerted vehicles, and reduced animal threat-perception that converts pre-crossing retreats into successful traversals (a 77% increase in safe road-crossing throughput). Sensitivity analyses establish a statistically significant equivalent-performance band across 5-20 m alternating radar spacing, with a recommended conservative 20 m operational deployment that coincides with the last statistically significant point in the sweep and preserves engineering margin against range-dependent radar SNR and environmental factors omitted from the simulation. Significant reductions are obtained across small-to-medium animal classes (fox- through deer-class targets) while reductions for elk- and moose-class animals trend downward and do not reach statistical significance at the trial count tested, reflecting the size-independent vehicle response in the simulator and consistent with the road-ecology finding that physical exclusion remains preferable for large-ungulate-dominated corridors. The system retains its collision-reduction effectiveness against tenfold degradation of baseline sensor sensitivity. The architecture is positioned at approximately one order of magnitude lower per-kilometre cost than comprehensive fence-and-crossing infrastructure and is appropriate for the substantial fraction of rural road kilometres for which dedicated fence-and-overpass solutions are economically out of reach. Field validation against an operational deployment is the necessary next step.

**Supplementary Materials:** The complete simulator source code (Python 3.x, MIT licence), analysis pipeline, raw per-trial CSV outputs for all experiments reported in this paper, and an interactive data exploration page with figures rendered as zoomable Plotly visualisations are available online (see the Data Availability Statement below).

**Author Contributions:** Conceptualization, L.T.; methodology, validation, resources, data curation, writing, original draft preparation, writing, review and editing, visualization. S.M. software, investigation, formal analysis, concept design, hardware design. All authors have read and agreed to the published version of the manuscript.

**Funding:** This research received no external funding.

**Institutional Review Board Statement:** Not applicable.

**Informed Consent Statement:** Not applicable.

**Data Availability Statement:** All code, raw data, and analysis pipelines required to reproduce the results presented in this paper are openly available at:
https://github.com/Gnacode/WILDLIFE-VEHICLE-COLLISION-MONTECARLO

and archived at Zenodo, release v1.1.0, DOI:
https://doi.org/10.5281/zenodo.20320115.

An interactive supplementary data exploration page is hosted at:
https://gnacode.github.io/WILDLIFE-VEHICLE-COLLISION-MONTECARLO/

**Acknowledgments:** This work draws on the GNACODE GNA radar-magnetometer platform originally developed for perimeter security applications.

**Conflicts of Interest:** L.T. is the founder and CEO of GNACODE Inc., which holds commercial interests in the radar-magnetometer sensor platform described in this paper. S.M. declares no conflict of interest.